\documentclass[conference]{IEEEtran}
\usepackage{cite}

\ifCLASSINFOpdf
   \usepackage[pdftex]{graphicx}
\else
   \usepackage[dvips]{graphicx}
\fi
\usepackage{amsmath}
\usepackage{fancyhdr}

\fancypagestyle{firstpage}
{
\pagenumbering{gobble}
\lfoot{978-1-5090-2760-6/17/\$31.00 \copyright 2017 IEEE}
}

\begin{document}
%
\title{Getting information from the\\ mixed electrical-heat noise}

\author{\IEEEauthorblockN{Adeline Cr\'epieux}
\IEEEauthorblockA{Aix Marseille Univ, Universit\'e de Toulon\\ CNRS, CPT, Marseille, France\\
Email: adeline.crepieux@cpt.univ-mrs.fr}
\and
\IEEEauthorblockN{Paul Eym\'eoud}
\IEEEauthorblockA{Aix Marseille Univ\\ CNRS, CINAM, Marseille, France}
\and
\IEEEauthorblockN{Fabienne Michelini}
\IEEEauthorblockA{Aix Marseille Univ, Univ Toulon\\ CNRS, IM2NP, Marseille, France }}


%


\maketitle
\thispagestyle{firstpage}

\begin{abstract}
We give a classification of the different types of noise in a quantum dot, for variable temperature, voltage and frequency. It allows us first to show which kind of information can be extracted from the electrical noise, such as the ac-conductance or the Fano factor. And next, to classify the mixed electrical-heat noise, and to identify in which regimes information on the Seebeck coefficient, on the thermoelectric figure of merit, or on the thermoelectric efficiency can be obtained.  
\end{abstract}


%
\IEEEpeerreviewmaketitle


\section{Introduction}

Since the pioneer works made by  Schottky~\cite{Schottky1918}, Johnson~\cite{Johnson1928} and Nyquist~\cite{Nyquist1928} hundred years ago, the current fluctuations in conductors have been studied without interruption both experimentally and theoretically. The results are abundant. Since then, the fluctuations are no always viewed as a trouble, or as an effect that has to be attenuated, but rather as a significant and accessible physical quantity which can bring information on the dynamics of charge carriers.

The objective here is not to give a historical panorama of the manifold results concerning the noise in nanoscale systems, since it exists several review articles on that topic~\cite{Rogovin1974,Kogan1996,Blanter2000,Cohen2005,Martin2005}. Instead, we propose a classification of the various types of noises according to the experimental conditions, such as the applied temperature $T$ and voltage $V$, as well as the measuring frequency $\omega$. We also discuss the weak transmission limit through the nanostructure. A particular attention will be devoted to the mixed noise, defined as the correlator between the electrical current and the heat current. Indeed, it has been  shown recently \cite{Crepieux2015,Eymeoud2016} that this quantity is relevant for the quantification of the thermoelectric conversion in quantum dot, which is the subject of an increasing number of experimental works \cite{Staring1993,Dzurak1998,Scheibner2005,Fahlvik2012,Fahlvik2013,Svilans2015,Svilans2016}.

The paper is organized as follows. In Sec. II, we recall the results obtained at equilibrium and out-of-equilibrium for the finite-frequency electrical noise and give its classification. In Sec. III, we do the same for the mixed noise and show how this quantity provides information on the thermoelectric conversion efficiency. We conclude in Sec. IV.

%

%
\section{Electrical Noise}

The noise we consider is the finite-frequency non-symmetrized noise \cite{Rothstein2009,Hammer2011,Zamoum2016} defined as: $\mathcal{S}_{\alpha\beta}^\mathrm{II}(\omega)=\int dt \exp(-i\omega t)\langle \Delta \hat I_\alpha(t) \Delta\hat I_\beta(0)\rangle$, where $\Delta \hat I_\alpha(t)=\hat I_\alpha(t)-\langle\hat I_\alpha\rangle$, with $\alpha$ the reservoir index ($\alpha=L$ for the left reservoir, and $\alpha=R$ for the right reservoir). For nanoscale systems, it is needed to make the distinction between the non-symmetrized noise and the symmetrized one for the reason that the electrical current operators, $\hat I_\alpha(t)$ and $\hat I_\beta(0)$, do not commute with each other \cite{Lesovik1997}. Besides, it is known that the non-symmetrized noise gives the emission noise at positive frequency, and the absorption noise at negative frequency \cite{Aguado2000}. Moreover, from the non-symmetrized noise, we can obtain the symmetrized one by simply taking $[\mathcal{S}_{\alpha\beta}^\mathrm{II}(\omega)+\mathcal{S}_{\beta\alpha}^\mathrm{II}(-\omega)]/2$. Experimentally, the measured noise depends on the way the detector works: when the detector is passive \cite{Lesovik1997,Gavish2000}, the measured noise is the non-symmetrized one at positive frequency (i.e., the emission noise), while for a classical detector, the measured noise is the symmetrized one \cite{Gabelli2009}. Here, we have chosen to focus on the emission noise. Note that the non-symmetrized noise and the symmetrized noise coincide in the limit of zero frequency, but also in the limit of high temperature.

\subsection{Johnson-Nyquist noise}

The Johnson-Nyquist noise corresponds to the noise observed in the limit of high temperature in comparison to the voltage and frequency: $k_BT\gg\{eV,\hbar\omega\}$. The system is at equilibrium and the fluctuation-dissipation theorem (FDT) holds. In that limit, the auto-correlator $\mathcal{S}_{\alpha\alpha}^\mathrm{II}(\omega)$ becomes independent of the reservoir index and is given by \cite{Johnson1928,Nyquist1928}
\begin{eqnarray}\label{JN}
 \mathcal{S}^\mathrm{II}(\omega)=2\hbar\omega N(\omega)G(\omega)~,
\end{eqnarray}
 where $N(\omega)=[\exp(\hbar\omega/k_BT)-1]^{-1}$ is the Bose-Einstein distribution function, and $G(\omega)$ the ac-conductance, i.e., the response of the system to an ac-voltage. Thus, the knowledge of the finite-frequency noise gives information on the dynamics of the carriers. If one goes to strictly zero-frequency, a direct expansion of the Bose-Einstein function leads to $\mathcal{S}^\mathrm{II}(0)=2k_BTG_\mathrm{dc}$, where $G_\mathrm{dc}$ is the dc-conductance. The zero-frequency noise gives information on the dc-conductance and can be used to determine the temperature of the system, as done for example in a quantum point contact \cite{Jezouin2013}. We call it thermal noise. It is also instructive to look at the symmetrized noise obtained from Eq.~(\ref{JN}). We get
\begin{eqnarray}\label{JNsym}
\mathcal{S}^\mathrm{II}_\mathrm{sym}(\omega)=2\hbar\omega \left[\frac{1}{2}+N(\omega)\right]G(\omega)~,
\end{eqnarray}
thanks to the even parity of the ac-conductance with frequency, and to the relation $N(-\omega)=-1-N(\omega)$. The factor 1/2 in Eq.~(\ref{JNsym}) corresponds to what is called zero-point noise fluctuations~\cite{Cohen2005}. It only appears when one considers the symmetrized noise.

\subsection{Cancellation of emission noise}

At voltage and frequency much higher than the temperature, $\{eV,\hbar\omega\}\gg k_BT$, the emission noise cancels when the frequency is higher than the applied voltage, due to the fact that when the detector is passive, the system can not emit energy larger than the one which is provided to it, here the voltage, thus we have $\mathcal{S}_{\alpha\beta}^\mathrm{II}(\omega>eV/\hbar)=0$. This property is specific to the non-symmetrized noise and has been confirmed experimentally in quantum dots \cite{Deblock2003,Basset2012,Delagrange2017} and in tunnel junctions~\cite{Parlavecchio2015,Fevrier2017}. It has been observed nonetheless that in some situations, as for example in the presence of electron-plasmon polariton interactions, the emitted signal can be extended to the region $\hbar\omega \in [eV,2eV]$ due to two-particles processes \cite{Schull2009,Xu2016}. One can wonder whether many-particles processes could induce a signal at $\hbar\omega>2eV$. However, this signal, if it exists, should be quite small.

\subsection{Out-of-equilibrium noise}

The out-of-equilibrium noise corresponds to the noise observed when there is no relation of order between the voltage, the frequency and the temperature. In that case, it has been shown \cite{Crepieux2016}  that we have for a quantum dot an equality between  the total ac-conductance and the anti-symmetrized noise, summed over reservoirs, which is given by
\begin{eqnarray}\label{OOE}
2\hbar\omega\sum_\alpha {G_\alpha}(\omega)=\sum_{\alpha,\beta}\Big[S^\mathrm{II}_{\alpha\beta}(-\omega)-S^\mathrm{II}_{\beta\alpha}(\omega)\Big]~.
\end{eqnarray}
This equality applies whatever are the values of frequency, voltage, temperature and transmission through the nanostructure. It can be viewed as a generalization of the FDT out-of-equilibrium. Note that previously a similar but non-equivalent relation was used, in which the sum over the reservoirs was absent~\cite{Safi2008,Safi2011,Roussel2016}. It is important to underline that out-of-equilibrium, there is no direct relation between the absorption noise $S_{\alpha\alpha}^\mathrm{II}(-\omega)$, and the emission noise $S_{\alpha\alpha}^\mathrm{II}(\omega)$. On the contrary, at equilibrium, thanks to the KMS relation \cite{Kubo1957,Martin1959}: $S_{\alpha\alpha}^\mathrm{II}(-\omega)=e^{\hbar\omega/k_BT}S_{\alpha\alpha}^\mathrm{II}(\omega)$, we find Eq. (\ref{JN}) back.

\subsection{Schottky noise}

At zero-frequency, when the transmission through the nanostructure is weak, the noise becomes proportional to the current $I=|\langle \hat I_\alpha\rangle|$, through the relation \cite{Schottky1918,Blanter2000} 
\begin{eqnarray}\label{Fano}
\mathcal{S}^\mathrm{II}_{\alpha\alpha}(0)=FeI\coth\left(\frac{eV}{2k_BT}\right)~,
\end{eqnarray}
which reduces to $\mathcal{S}^\mathrm{II}_{\alpha\alpha}(0)=FeI$ at low temperature. Note that because of charge conservation, we have in case of two-terminal nanosystem: $|\langle \hat I_L\rangle|=|\langle \hat I_R\rangle|$. The Fano factor $F$ takes various values according to the nature of the system: $F=1$ for metallic reservoirs, $F=2$ for superconducting reservoirs, $F=1/3$ for a quantum point contact working in the fractional quantum Hall regime \cite{Saminadayar1997Picciotto1997}, $F=(1+g^2)/2$ in a carbon nanotube in which electrons are injected from a STM tip \cite{Crepieux2003}, where $g<1$ is the Coulomb interaction parameter, and $F=5/3$ for a Kondo quantum dot \cite{Sela2006,Zarchin2008,Ferrier2016}.

At finite-frequency, weak energy independent transmission, and $F=1$, the noise is equal to the sum of two terms involving the current $I$ at shifted voltages \cite{Rogovin1974,Fevrier2017,Roussel2016}, that is
\begin{eqnarray}
 \mathcal{S}^\mathrm{II}_{\alpha\alpha}(\omega)=e\sum_\pm N(\omega\pm eV/\hbar)I(\hbar\omega/e \pm V)~,
\end{eqnarray}
which is fully consistent with Eq.~(\ref{Fano}) in the limit of zero frequency, and with Eq.~(\ref{JN}) in the limit of zero voltage, provided that $I(\omega)=\hbar\omega G(\omega)/e$.

\subsection{Classification of emission noises}

\begin{figure}[!h]
\centering
\includegraphics[width=8.5cm]{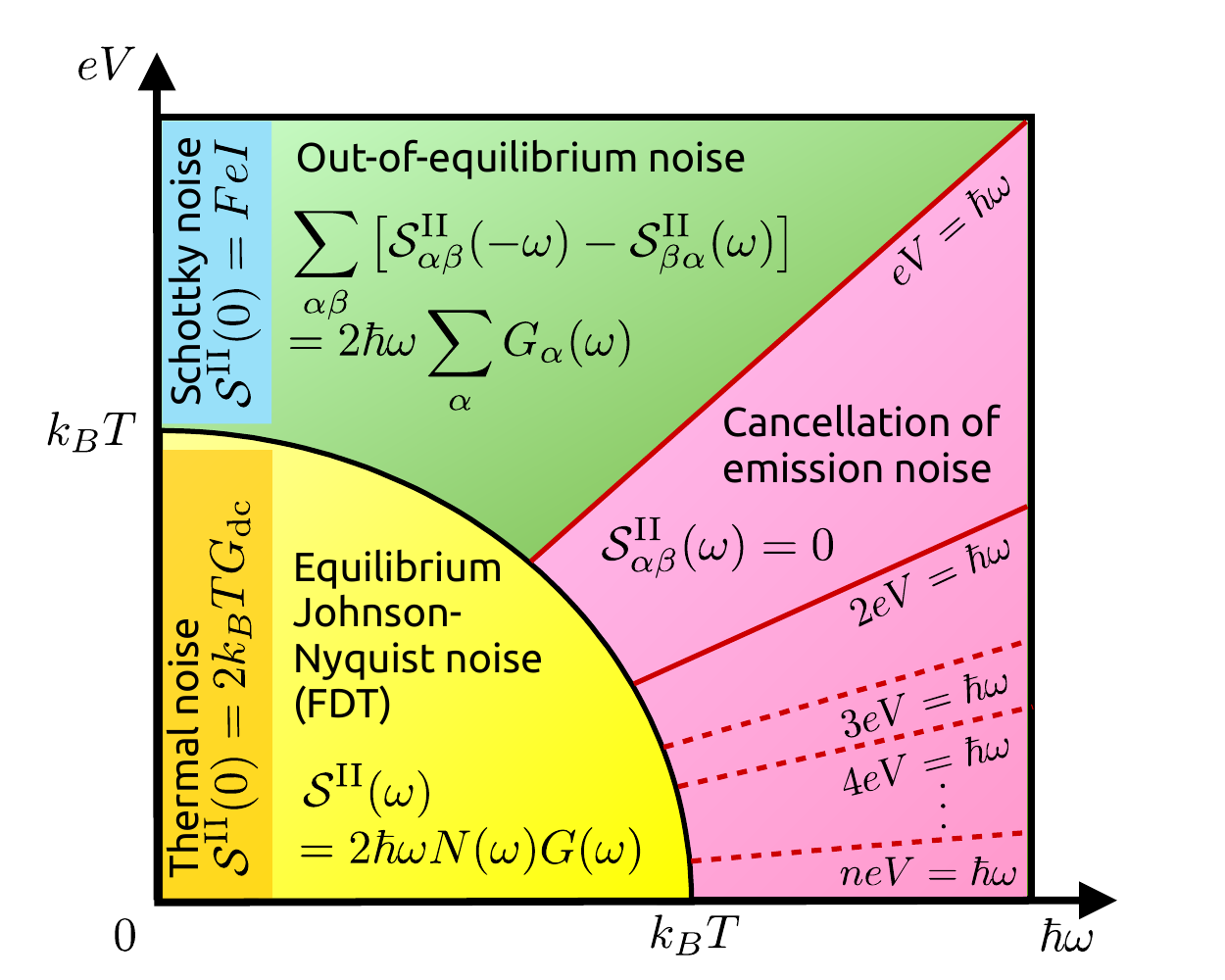}
\vspace*{-0.2cm}
\caption{Electrical noise phase diagram associated to a quantum dot.}
\label{noise}
\end{figure}

The set of behaviors described in Secs. II.A to II.D are summarized in Fig.~\ref{noise}. The yellow region corresponds to the region where the system is at equilibrium since $k_BT\gg\{eV,\hbar\omega\}$: the FDT, given by Eq.~(\ref{JN}) of Sec. II.A, applies and the noise corresponds to the Johnson-Nyquist noise. At zero frequency, the noise reduces to the thermal noise only (orange region). The pink region is the region where the emission noise cancels as explained in Sec. II.B. A signal can eventually be measured in the region between the solid red straight lines when interactions are present, like in Ref.~\cite{Schull2009}. The green region is the region of out-of-equilibrium noise where Eq.~(\ref{OOE}), given in Sec.~II.C, applies. This relation remains to be tested experimentally. At zero frequency and  for weak transmission, the noise reduces to the Schottky noise (also called shot noise) and becomes proportional to the current, as explained in Sec.~II.D (blue region). Thus, according to the region in which the noise measurement is made, it gives access either to the Fano factor, either to the dc-conductance or to the ac-conductance. It can also indicate the presence of interactions if a non-zero signal is observed in the pink region.


\section{Mixed Noise}

The finite-frequency non-symmetrized mixed noise is defined as the Fourier transform of the correlator between the electrical current operator, $\hat I_\alpha$, and the heat current operator, $\hat J_\beta$, such as: $\mathcal{S}_{\alpha\beta}^\mathrm{IJ}(\omega)=\int dt \exp(-i\omega t)\langle \Delta \hat I_\alpha(t) \Delta\hat J_\beta(0)\rangle$, where  $\Delta \hat J_\beta(t)=\hat  J_\beta(t)-\langle\hat  J_\beta\rangle$. Even if there are up to now only few works devoted to this quantity, we show below that many information can be extracted from it, in particular concerning the thermoelectric conversion. It makes this quantity unavoidable and worthy of interest.

\subsection{Equilibrium mixed noise}

It has been shown recently~\cite{Crepieux2016} that at temperature much higher than the voltage and frequency, $k_BT\gg\{eV,\hbar\omega\}$, the mixed noise reads as
\begin{eqnarray}\label{SIJeq}
\mathrm{Re}\Big\{\sum_{\alpha\beta}\mathcal{S}^\mathrm{IJ}_{\alpha\beta}(\omega)\Big\}=2\hbar\omega N(\omega)\sum_\alpha G_\alpha^\mathrm{T}(\omega)~,
\end{eqnarray}
where $G_\alpha^\mathrm{T}(\omega)$ is the thermoelectric ac-conductance, i.e., the real part of the thermoelectric admittance, defined as the derivative of the first harmonic of the heat current according to the ac-gate voltage. It is related to the Seebeck ac-coefficient, $S^T_\alpha(\omega)$, and to the electrical ac-conductance, $G_\alpha(\omega)$, through the relation: $G_\alpha^\mathrm{T}(\omega)=TS^T_\alpha(\omega)G_\alpha(\omega)$. In the zero-frequency limit, Eq.~(\ref{SIJeq}) reduces to
\begin{eqnarray}\label{SST}
\mathcal{S}^\mathrm{IJ}(0)&=&2k_BT^2S_\mathrm{dc}^TG_\mathrm{dc}~,
\end{eqnarray}
where $S_\mathrm{dc}^T$ is the Seebeck dc-coefficient. The reservoir indices are omitted to shorten the notations. In addition we have $\mathcal{S}^\mathrm{JJ}(0)=2k_BT^2[K_\mathrm{dc}+(S_\mathrm{dc})^2TG_\mathrm{dc}]$, where $K_\mathrm{dc}$ is the thermal dc-conductance. $\mathcal{S}^\mathrm{JJ}(0)$ corresponds to the heat noise at zero-frequency, i.e., the correlator between the heat current and itself. Using these expressions, together with the relation $\mathcal{S}^\mathrm{II}(0)=2k_BTG_\mathrm{dc}$ given in Sec.~II.A, the thermoelectric figure of merit, defined as $ZT=(S_\mathrm{dc}^T)^2TG_\mathrm{dc}/K_\mathrm{dc}$, can be written equivalently under the form \cite{Crepieux2015}
\begin{eqnarray}\label{ZT}
ZT=\frac{[\mathcal{S}^\mathrm{IJ}(0)]^2}{\mathcal{S}^\mathrm{II}(0)S^\mathrm{JJ}(0)-[\mathcal{S}^\mathrm{IJ}(0)]^2}~.
\end{eqnarray}
Note that $ZT$ is positive thanks to Cauchy-Schwarz inequality which ensures a positive denominator. Thus, at equilibrium, the figure of merit can be fully expressed in terms of electrical, mixed and heat noises, and it cancels when the mixed noise cancels, meaning that the mixed noise is a key witness of the thermoelectric conversion occurring in the system.

\subsection{Cancellation of mixed noise}

Similarly to what happens for the emission noise (see Sec.~II.B),  the mixed noise cancels at frequency higher than voltage provided that temperature is low, thus we have: $\mathcal{S}_{\alpha\beta}^\mathrm{IJ}(\omega>eV/\hbar)=0$. It has been proved for a non-interacting quantum dot \cite{Eymeoud2016}. The effect of interactions on the mixed noise has not yet been studied, but we can conjecture that, as it is the case for the electrical noise, the mixed noise could be non-zero in the region $\hbar\omega>eV$ due to many-particles processes.

\subsection{Out-of-equilibrium mixed noise}

When the voltage, the frequency and the temperature are all of the same order of magnitude, there is no simple relation between the mixed noise and the thermoelectric ac-conductance, opposite to what we had for the electrical noise and the ac-conductance which are related together through Eq.~(\ref{OOE}). However, for frequency much higher than temperature, such a relation exists \cite{Crepieux2016}, and takes the form
\begin{eqnarray}\label{SIJ_OOE}
2\hbar\omega G_\alpha^T(\omega)=\mathrm{Re}\left\{S^\mathrm{IJ}_{\alpha\alpha}(-\omega)-S^\mathrm{JI}_{\alpha\alpha}(\omega)\right\}~,
\end{eqnarray}
which means that the asymmetry of the mixed noise is related to the thermoelectric ac-conductance, itself related to the Seebeck ac-coefficient.

\subsection{Schottky mixed noise}

At zero-frequency and zero-temperature, and in the limit of weak transmission through the quantum dot, assumed to be non-interacting, the mixed noise is proportional to the heat current, $J_\alpha=\langle \hat J_\alpha\rangle$, in the same manner that the electrical noise is proportional to the electrical current $I$ (see Sec.~II.D). Indeed, we have at low temperature~\cite{Crepieux2015}
\begin{eqnarray}
\mathcal{S}^\mathrm{II}_\mathrm{LR}(0)&=&-eI~,\nonumber\\\label{SIJ}
\mathcal{S}^\mathrm{IJ}_\mathrm{LR}(0)&=&eJ_R=-(\varepsilon_0-\mu_R)I~,\\
\mathcal{S}^\mathrm{JJ}_\mathrm{LR}(0)&=&(\varepsilon_0-\mu_L)J_R~,\nonumber
\end{eqnarray}
where $\mu_\mathrm{L,R}$ are the chemical potentials of the left and right reservoirs, and $\varepsilon_0$, the energy of the quantum dot which is assumed to have only one level. Note that we have $|J_L|\ne |J_R|$ since the heat, contrary to charge, is not a conserved quantity.
The above relations allow us to rewrite the thermoelectric efficiency, defined as $\eta=|J_R/VI|$ in case of an electricity to heat conversion, in terms of noises. Indeed, we have \cite{Crepieux2015}
\begin{eqnarray}\label{eta}
\eta=\frac{[\mathcal{S}^\mathrm{IJ}_\mathrm{LR}(0)]^2}{\left|\mathcal{S}^\mathrm{II}_\mathrm{LR}(0)\mathcal{S}^\mathrm{JJ}_\mathrm{LR}(0)-[\mathcal{S}^\mathrm{IJ}_\mathrm{LR}(0)]^2\right|}~.
\end{eqnarray}
Thus, in the limit of weak transmission, the thermoelectric efficiency can be fully expressed in terms of electrical, mixed and heat noises, and it cancels when the mixed noise cancels, meaning once again that the mixed noise is a key witness of the thermoelectric conversion, as was expected intuitively.

At finite-frequency and weak energy independent transmission, the mixed noise is equal to the sum of two terms involving the heat current $J_\beta$ at shifted voltages, that is
\begin{eqnarray}
 \mathcal{S}^\mathrm{IJ}_{\alpha\beta}(\omega)=e\sum_\pm \pm N(\omega\pm eV/\hbar)J_\beta(\hbar\omega/e \pm V)~,
\end{eqnarray}
which is fully consistent with Eq.~(\ref{SIJ}) in the limits of zero-frequency and temperature.

\subsection{Classification of mixed noises}

\begin{figure}[!h]
\centering
\includegraphics[width=8cm]{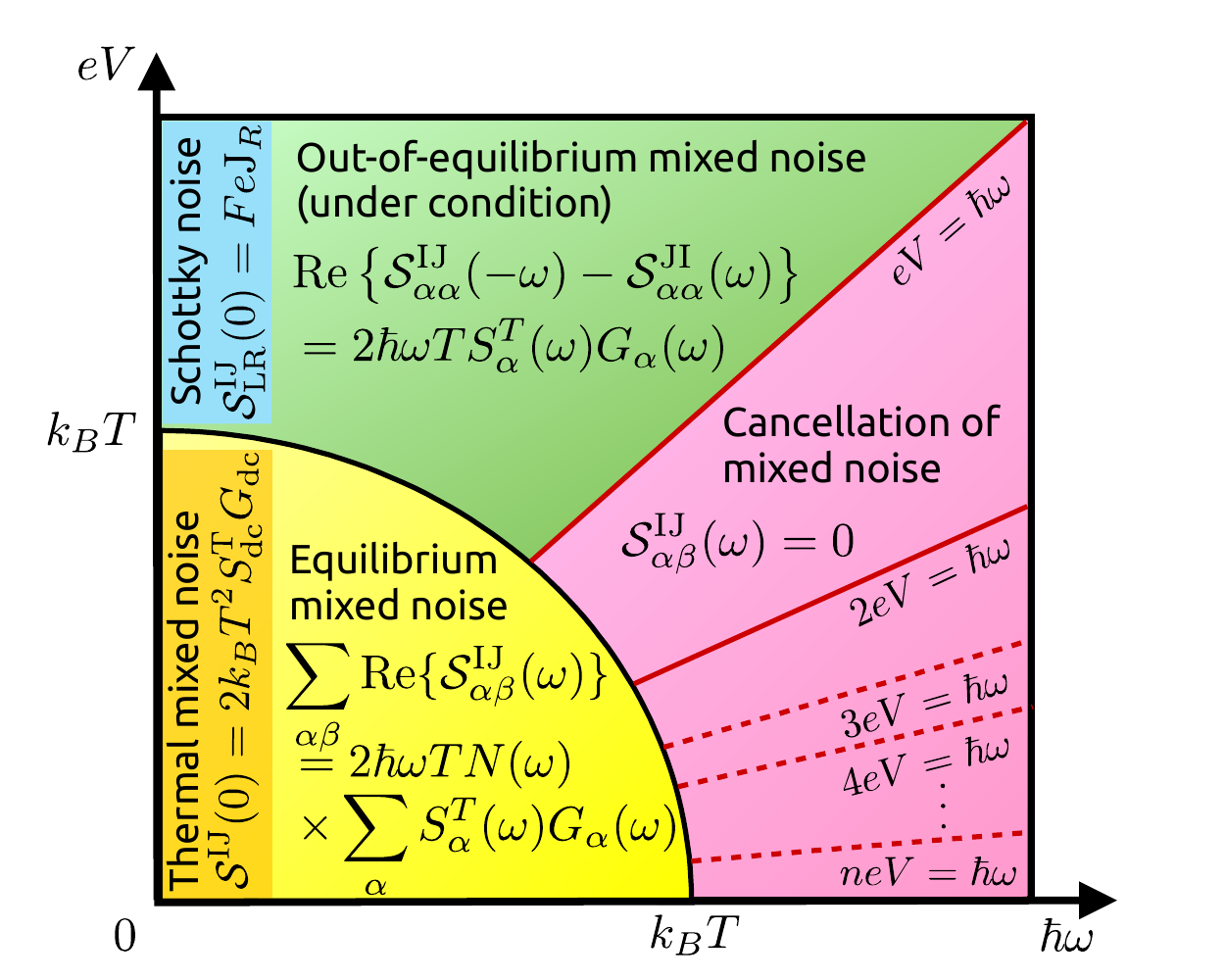}
\vspace*{-0.3cm}
\caption{Mixed noise phase diagram associated to a quantum dot.}
\label{mixed}
\end{figure}

The set of behaviors described in Secs. III.A to III.D are summarized in Fig.~\ref{mixed}. The yellow region corresponds to the region where the system is at equilibrium since $k_BT\gg\{eV,\hbar\omega\}$, the FDT generalized for mixed noise and given by Eq.~(\ref{SIJeq}) of Sec.~III.A, applies. At zero frequency (orange region), the mixed noise becomes proportional to the Seebeck coefficient, $S^T_\mathrm{dc}$, as stated by Eq.~(\ref{SST}). The pink region is the region where the mixed noise cancels as explained in Sec.~III.B. The green region is the region of out-of-equilibrium noise where Eq.~(\ref{SIJ_OOE}), given in Sec.~III.C, applies under conditions, as detailed above. At zero frequency, the mixed noise reduces to the Schottky noise and becomes proportional to the heat current for weak transmission, as explained in Sec.~III.D (blue region). Thus, according to the region inside which the mixed noise is considered, the knowledge of this quantity gives access either to the Seebeck coefficient, either to the heat current, or to the thermoelectric ac-conductance. Consequently, the ratio of noises $[\mathcal{S}^\mathrm{IJ}_\mathrm{LR}(0)]^2/|\mathcal{S}^\mathrm{II}_\mathrm{LR}(0)\mathcal{S}^\mathrm{JJ}_\mathrm{LR}(0)-[\mathcal{S}^\mathrm{IJ}_\mathrm{LR}(0)]^2|$ gives the thermoelectric figure of merit in the orange region, and the thermoelectric efficiency in the blue region. 

\vspace*{-0.3cm}

\vspace*{-0.3cm}

\section{Conclusion}
Varying the temperature, the voltage and the frequency, we have highlighted the rich behavior of the non-symmetrized electrical and mixed noises in a quantum dot. With the phase diagram of Fig.~\ref{noise}, we have identified the condition to obtain either out-of-equilibrium noise, equilibrium noise, thermal noise, shot noise or vanishing noise. In addition of giving a global picture of what information can be extracted from the mixed noise, such as the thermoelectric figure of merit or the thermoelectric efficiency, the phase diagram of Fig.~\ref{mixed} has allowed us to identify the issues which remain to be studied. These open issues are:  (i) does the Fano factor appear in the expression of Schottky mixed noise, (ii) do the interactions affect the cancellation of mixed noise at frequency larger than voltage, (iii) do the $n$-particles processes lead to non-zero signal in the regions $\hbar\omega>neV$ (delimited by the solid and dashed red lines in Figs.~1 and 2), and (iv) is the FDT generalized out-of-equilibrium verified experimentally for both electrical and mixed noises?


\section*{Acknowledgment}

The authors would like to thank W.~Belzig, R.~Deblock, J.~Gabelli, M.~Guigou, M.~Lavagna, T.~Martin, F.~Portier, P.~Samuelsson, and R.~Zamoum for discussions.




%

\end{document}